\documentclass
[aps,prd,twocolumn,floats,balancelastpage,showpacs,preprintnumbers,prd,amssymb,floatfix,nofootinbib,amsmath]{revtex4}

\usepackage{epsfig}
\usepackage{epstopdf}

\usepackage{float}
\usepackage[usenames]{color}
\usepackage{color,epsfig,amsmath,amssymb,fancyhdr}   
\usepackage[french]{babel}
\usepackage{ulem}   

\begin{document}
\preprint{IRFU-12-65}

\title{Irregularity in gamma ray source spectra as a signature of axionlike particles}

\author{Denis Wouters}
 \email{denis.wouters@cea.fr}
\affiliation{CEA, Irfu, Centre de Saclay, F-91191
Gif-sur-Yvette | France}

\author{Pierre Brun}
\email{pierre.brun@cea.fr}
\affiliation{CEA, Irfu, Centre de Saclay, F-91191
Gif-sur-Yvette | France}

\begin{abstract}

Oscillations from high energy photons into light pseudoscalar particles in an external magnetic field are expected to occur in some extensions of the standard model. It is usually assumed that those axionlike particles (ALPs) could produce a drop in the energy spectra of gamma ray sources and possibly decrease the opacity of the Universe for TeV gamma rays. We show here that these assumptions are in fact based on an average behavior that cannot happen in real observations of single sources. We propose a new method to search for photon-ALP oscillations, taking advantage of the fact that a single observation would deviate from the average expectation. Our method is based on the search for irregularities in the energy spectra of gamma ray sources.  We predict features that are unlikely to be produced by known astrophysical processes and a new signature of ALPs that is easily falsifiable.
\end{abstract}


\maketitle

\section{Introduction}

Light pseudoscalar particles  appear in many extensions of the standard model. The most typical example is the axion, which was introduced as a consequence of the Peccei-Quinn mechanism to solve the puzzle of the absence of CP violation in quantum chromodynamics~\cite{Peccei:1977hh, Peccei:1977ur}. The axion is a hypothetical light particle that has a two-photon vertex described by the interaction term
\begin{equation}
\mathcal{L}_{a\gamma}\;\;=\;\;-\frac{1}{4}gF_{\mu\nu}\tilde{F}^{\mu\nu}a\;\; = \;\;g\;\vec{E}\cdot\vec{B}\;a\;\;,
\end{equation}
where $g$ is the axion-photon coupling constant, $F$ is the electromagnetic tensor, $\tilde{F}$ its dual, $\vec{E}$ the electric field, $\vec{B}$ the magnetic field and $a$ the axion field. This term implies the possibility of photon-axion oscillations in an external magnetic field~\cite{1983PhRvL..51.1415S, 1988PhRvD..37.1237R}. This coupling is used experimentally to search for axions that would be thermally produced in the Sun~\cite{2011PhRvL.107z1302A}, or axion dark matter~\cite{2010PhRvL.105q1801W}. In the case of the Peccei-Quinn axion, the photon-axion coupling is predicted to scale with the axion mass; however, other models predict light pseudoscalar particles with the same coupling to the electromagnetic field but {\it a priori} unrelated to their mass~\cite{1987PhR...150....1K}. Those are called axionlike particles (ALPs), the phenomenology of which is similar to standard axions. Astrophysical environments can offer ideal conditions for photon-ALP oscillations, with the possibility of long baseline experiments involving magnetic fields~\cite{1996slfp.book.....R}. Progress over the last decade in $\gamma$-ray astronomy allowed one to consider searching for the imprints of $\gamma$-ALP oscillations in the energy spectra of high energy $\gamma$-ray sources~\cite{2007PhRvD..76b3001M}. The effect of $\gamma$-ALP oscillation is usually assumed to be twofold: it is expected to induce a dimming of the fluxes above a given threshold ~\cite{2007PhRvL..99w1102H, 2007PhRvD..76l3011H}, and possibly decrease the gamma-ray pair production related opacity at high energy. The opacity can be that of the intergalactic medium~\cite{2007PhRvD..76l1301D, 2009PhRvD..79l3511S} or within the sources themselves~\cite{2012arXiv1202.6529T}. A crucial point is the turbulent nature of the magnetic fields the photon beam travels through. It implies a consequential randomness in the prediction of the observable effects. This has been pointed out in~\cite{2009JCAP...12..004M} in the case of the change of opacity due to $\gamma$-ALP oscillations. The authors of~\cite{2009JCAP...12..004M, 2009PhRvL.102t1101B} showed that because of the random nature of the intergalactic magnetic fields, the effect of $\gamma$-ALP mixing should be very different from one source to another. Such an observable is then useless to perform ALP searches through the observation of a single source, leaving only the possibility of a population study in order to average the effect over many sources. This type of study has been conducted in {\it e.g.}~\cite{Horns:2012fx}, showing a hint for an anomaly in the transparency of the Universe. Though rapidly increasing with the advent of the last generation Cherenkov telescope arrays such as HESS, MAGIC and VERITAS, there are only a handful of high energy sources that are effectively concerned with extragalactic absorption. It is thus interesting to point out some effect of the $\gamma$-ALP mixing that does not rely either on stacking or averaging, in order to exploit observations of single sources. Here for the first time an effect is pointed out that potentially applies to single observations. This article is organized as follows. First, we briefly recall the formalism of $\gamma$-ALP mixing and apply the results to a single coherent magnetic domain. As a second step we show the results of a simulation of photons traveling through a set of magnetic domains. In particular we show that contrary to what is stated in the literature, a sharp drop in the energy spectrum of high-energy $\gamma$-ray sources is not a robust observable and is not what should be searched for. Actually the $\gamma$-ALP mixing would produce an anomalous dispersion of the spectra, which would no longer appear as smooth in a limited energy range. We then give an explicit example of how the effect could appear in the data, in the case of a specific situation, namely an extragalactic TeV emitter whose photons travel through the intergalactic magnetic field (IGMF), and we discuss the robustness of the method. 

\section{The photon/axion system in a magnetic field}

\begin{figure}[t]
\centering
\includegraphics[width=\columnwidth]{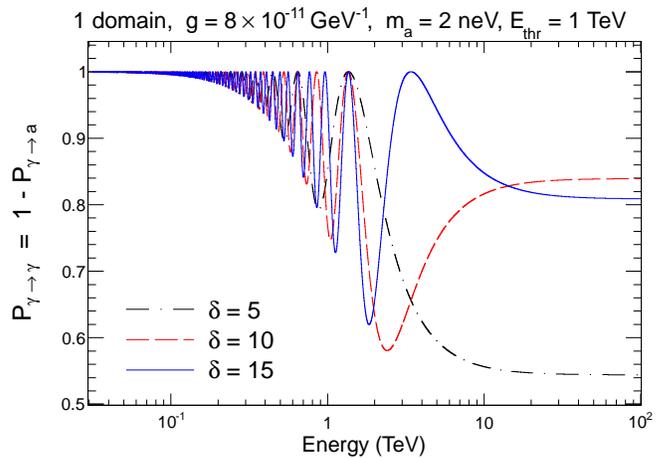}
\caption{Survival probability of an unpolarized photon as a function of the energy for three values of $\delta$.}
\label{fig:1domain}
\end{figure}

The $\gamma$-ALP system is described following the approach of~\cite{1988PhRvD..37.1237R}. A three-state wave function is used with two states of polarization for the photon and one for the ALP. Let $\theta$ be the angle between the magnetic field direction and the photon momentum. Since only the $\vec{B}$ component transverse to the propagation couples photons and ALPs, the strength of the magnetic field involved in the coupling is $B\sin\theta$. Moreover, for parity issues, only one polarization state parallel to the field is involved in the interaction. This is accounted for by introducing the angle $\phi$ between the transverse component of the field and the direction of the polarization sate $A_1$. The $\gamma$-ALP system is then propagated using the following linearized equations of motion assuming relativistic axions:
\begin{equation}
\left(E - i\partial_z + \mathcal{M} \right )
\left(\begin{array}{c} A_1 \\ A_2 \\ a \end{array} \right) = 0 \;\;,
\end{equation}
with the mixing matrix
\begin{equation}
\mathcal{M}\;\;=\;\;\left(\begin{array}{ccc} \Delta_{11}-i\Delta_{\mathrm{abs}} & \Delta_{12} & \Delta_\mathrm{B}\cos\phi \\ \Delta_{21} & \Delta_{22}-i\Delta_{\mathrm{abs}} & \Delta_\mathrm{B}\sin\phi \\ \Delta_\mathrm{B}\cos\phi & \Delta_\mathrm{B}\sin\phi\ & \Delta_\mathrm{a} \end{array}\right)\;\;,
\end{equation}
where $\Delta_\mathrm{B} = gB\sin\theta/2$  is the coupling term, and $\Delta_\mathrm{a} = -m_\mathrm{a}^2/2E$ is the ALP mass term. Here we neglect the Faraday effect and the vacuum Cotton-Mouton term, as the low magnetic field strength considered in the following makes the corresponding contribution irrelevant for this study. This implies $\Delta_{12} = \Delta_{21} = 0$ and that the other diagonal terms are $\Delta_{11} = \Delta_{22} = -\omega^2_\mathrm{pl}/2E$, $\omega_{\rm pl}$ being the plasma frequency accounting for the effective photon mass. As in~\cite{2003JCAP...05..005C}, absorption of photons on their way is introduced with the $\Delta_{\mathrm{abs}} = \tau/2s$ term where $\tau$ is the optical depth assuming a propagation over a domain of size $s$ within which the opacity is homogeneous. Because of that term, the matrix is no longer Hermitian and unitarity is lost. In the following, this term will be used to model the absorption of photons on the extragalactic background light (EBL) while propagating in IGMFs. After diagonalization of the mixing matrix, the equations of motion can be analytically solved and the transfer matrix of the system is obtained. The probability of $\gamma-a$ conversion after crossing one coherent magnetic field domain of size $s$ in the simplest case,  without absorption and neglecting the plasma term, yields
\begin{equation}\label{prob}
P_{\gamma\rightarrow a} \;=\; 
\frac{2 \Delta_\mathrm{B}^2}{\Delta_{\mathrm{osc}}^2}\sin^2\frac{\Delta_{\mathrm{osc}}s}{2}\;\;,
\end{equation}
with $\Delta_{\mathrm{osc}} = \sqrt{\Delta_\mathrm{a}^2+4\Delta_\mathrm{B}^2}$. The energy dependence of the mass terms in $\Delta_{\mathrm{osc}}$ implies an energy threshold above which the conversion becomes efficient, 
\begin{equation}\label{thresh}
E_\mathrm{thr} = \frac{m_{\mathrm{eff}}^2}{2gB\sin\theta} \;\; , 
\end{equation}
$m_\mathrm{eff}$ being the effective ALP mass in the presence of charges ({\it e.g.} in a plasma). For $E \ll E_\mathrm{thr}$, $\Delta_{\mathrm{osc}} \gg \Delta_\mathrm{B}$ and then no conversion occurs. For $E \sim E_{\mathrm{thr}}$ spectral oscillations happen due to the energy dependent $\sin^2\Delta_{\mathrm{osc}}s/2$ term. For $E \gg E_\mathrm{thr}$, $\Delta_{\mathrm{osc}} \sim \Delta_\mathrm{B}$ and the conversion probability is no longer energy dependent. The conversion probability of Eq.~\ref{prob} can be parameterized in terms of $E_\mathrm{thr}$ and
\begin{equation}\label{delta}
\delta = gBs\sin\theta/2
\end{equation}
instead of $B\sin\theta$ and $s$. $\sin^2\delta/2$ is then the conversion probability at very high energy (VHE, $E \gg E_\mathrm{thr}$).  The condition required for a significant conversion to occur, $\delta \gtrsim 1$, is similar to the Hillas criterion for the acceleration of ultra high energy cosmic rays, as pointed out in~\cite{2007PhRvL..99w1102H}. Figure ~\ref{fig:1domain} shows the evolution of the photon survival probability as function of the energy for three different values of $\delta$. For allowed large IGMF values of order 1 nG, an ALP mass of 2 neV and a coupling $g=8\times10^{-11}\;\rm GeV^{-1}$ at the limit of current experimental constraints~\cite{2011PhRvL.107z1302A}, $E_\mathrm{thr}$ lies at about 1 TeV.
The asymptotical value of $1-P_{\gamma \rightarrow a}$ gives the level of dimming of the photon flux (independently of eventual additional EBL absorption). One can see that this attenuation is hardly predictable given the uncertainties on the environmental parameters, $\vec{B}$ and $s$, as it depends sinusoidally on the value of $\delta$. 

\begin{figure}[t]
\centering
\includegraphics[width=\columnwidth]{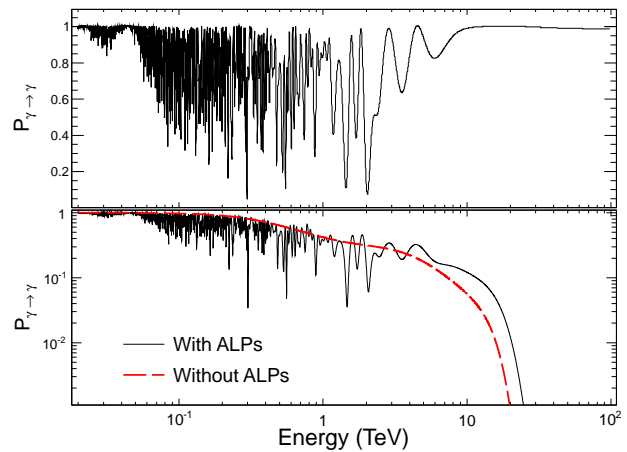}
\caption{Photon survival probability as function of the energy for a realization of a source at $z = 0.1$ using B = 1 nG, $s_0$ = 1 Mpc,  $g=8\times10^{-11}$ GeV$^{-1}$ and $m_{\rm a}$ = 2 neV without absorption (upper panel) and with EBL absorption (lower panel).
}
\label{fig:turbulent}
\end{figure}

In astrophysical environments, magnetic fields are usually not coherent. In the case of a propagation through a turbulent magnetic field, the beam path can be divided into coherent domains of size of the coherence length of the field (the validity of this simple model is discussed in Sec.~\ref{discussion}). For each domain, a transfer matrix is generated with a random orientation of the magnetic field yielding a specific value of $\delta$. The total transfer matrix associated with this realization of the turbulent magnetic field is the product of all individual transfer matrices. The spectral shape of the global conversion probability for one single realization is the result of the interference of all oscillation patterns such as those displayed in Fig.~\ref{fig:1domain}. As the pseudo-period is different in each domain, the photon survival probability has a very complex energy dependence. As an illustration, the survival probability of a photon from a source at redshift $z=0.1$ traveling through a single realization of a 1 nG IGMF with coherent domains of size $s_0=1\;\rm Mpc$ is displayed in Fig.~\ref{fig:turbulent}. A plasma density of $n_e = 10^{-7} \mathrm{cm}^{-3}$ typical of the intergalactic medium is assumed. In this condition and for ALP masses of order neV, $m_{\rm eff}=m_{\rm a}$. For illustration, the upper panel shows the survival probability without absorption on the EBL, whereas the lower panel results include this effect. Conservatively, the EBL density model used here is the lower limit model from~\cite{2010A&A...515A..19K}. To account for redshifting, a flat $\rm \Lambda CDM$ Universe with $(\Omega_{\rm \Lambda},\,\Omega_{\rm m})=(0.73,\,0.27)$ and $H_0=71\,\rm km/s/Mpc$ is assumed. Here the dashed red line is the prediction without ALPs, so that the dimming is only due to EBL. From Fig.~\ref{fig:turbulent} one can see that the prediction of the model including ALPs is the presence of a significant level of noise in the energy spectrum over one decade or so around $E_{\rm thr}$. Because of the unknown nature of the orientation of the magnetic field within the domains, the exact shape of the spectrum in this region is unpredictable. However, as we shall see in the following, the noise level is a prediction of the model. This prediction significantly differs from what usually appears in the literature, namely a smooth transition between no dimming below $E_{\rm thr}$ and a fixed level of attenuation above it. It has been shown in~\cite{2002PhLB..543...23G} that the averaging over a large number of realizations of  $N$ domains in each of which the conversion probability is $P_0$ yields an overall conversion probability
%
\begin{equation}\label{paveraged}
P_{\gamma\rightarrow a} =  \frac{1}{3}\left (1-{\rm e}^{-3NP_0} \right )\;\; .
\end{equation}
This means that the effect as it has been studied so far is valid for an average over a collection of sources. In the case of the observation of one source only, if $N$ is very large and the energy spectrum is binned, then the smooth behavior can be retrieved in principle. In practice $N$ is not large enough, as we shall see in the following. The results presented in Fig.~\ref{fig:turbulent} are obtained with a single realization. By averaging the results of Fig.~\ref{fig:turbulent} over a large number of realizations, the value given by Eq.~\ref{paveraged} is retrieved. From one realization to another, only the orientations of the magnetic fields vary; the number of domain and their sizes are kept fixed.

Note that above 5 TeV the survival probability for this specific realization is higher with ALPs than with EBL only. This is the so-called opacity effect, because photons are untouched by the EBL as they travel disguised as axions, the Universe appears to be more transparent. This result should be taken with care, however because, as shown in~\cite{2009JCAP...12..004M}, there exist realizations of the IGMF where the opposite effect is obtained, basically when most ALPs do not convert back to photons before detection. 

\section{Observational effects}\label{observational}

\begin{figure*}[t]
\centering
\includegraphics[width=.75\columnwidth]{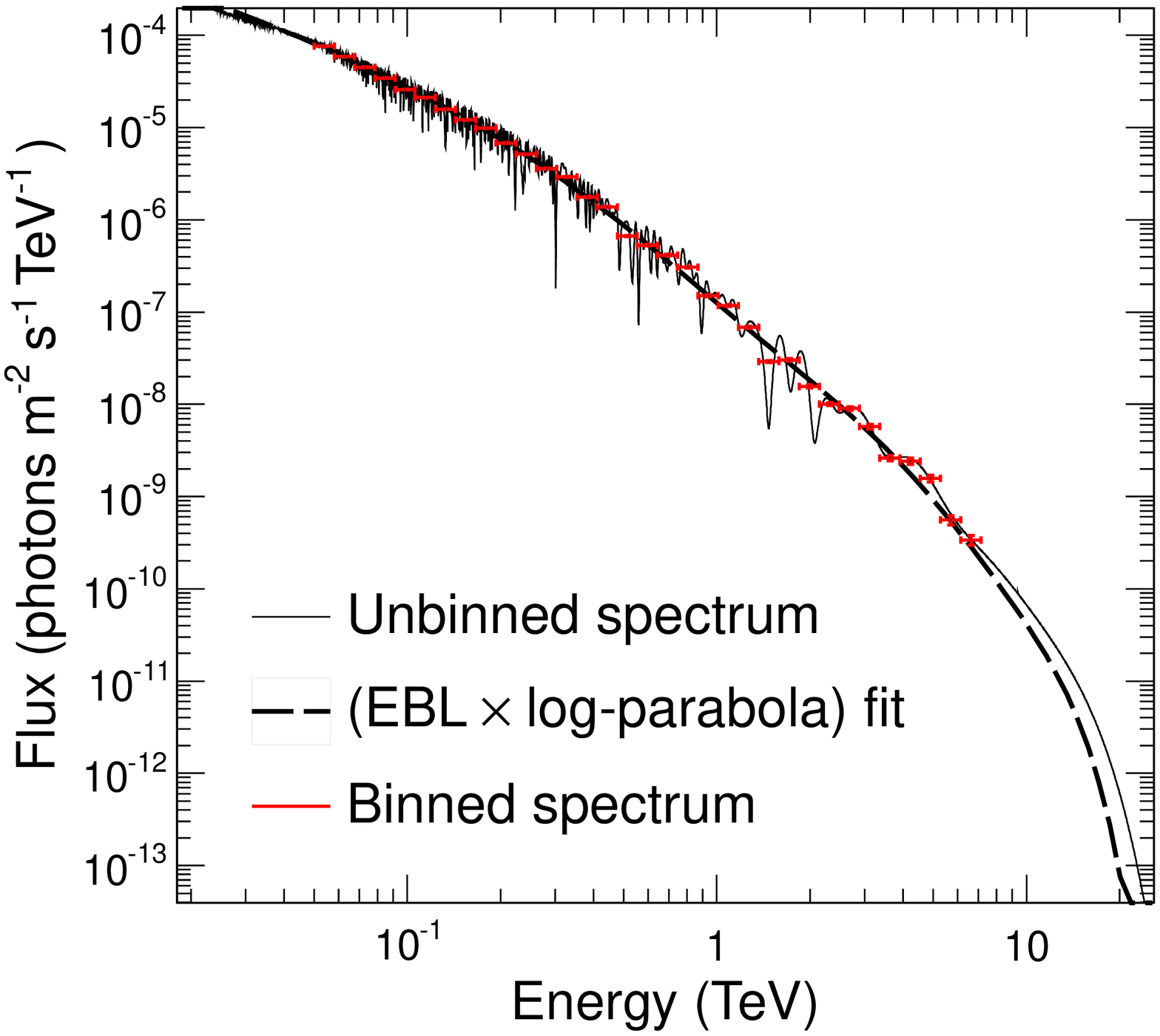}
\includegraphics[width=.75\columnwidth]{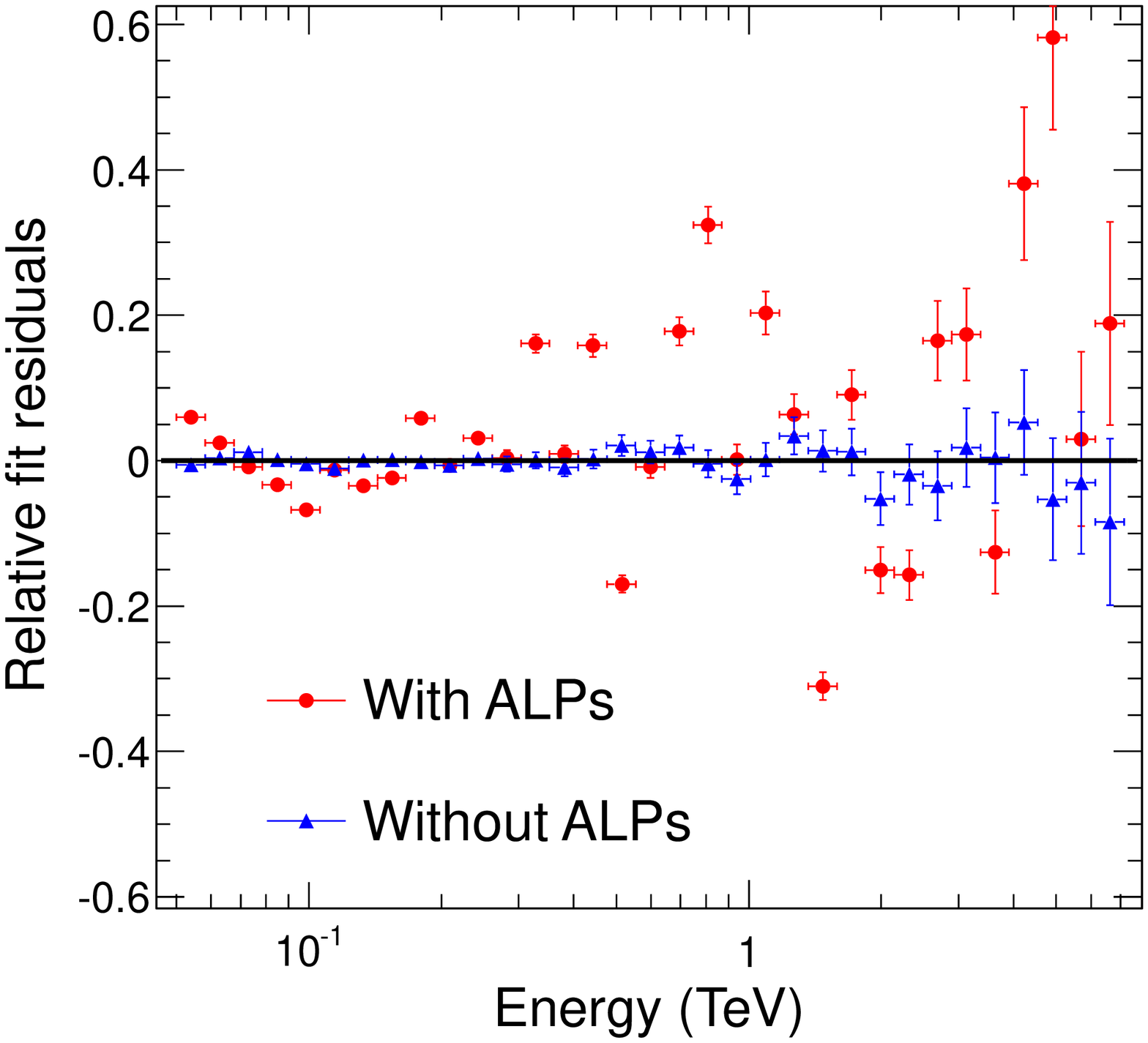}
\caption{Simulation of the observation of one $\gamma$-ray source at $z$=0.1, with the effect of $\gamma$-ALP mixing (left), and distribution of the residuals of fits to a conventional model and a model with ALPs (right).}
\label{fig:simu}
\end{figure*}

The experimental relevance of the proposed signature is now studied in the particular case of a source at redshift $z = 0.1$ for the same parameters as above. The intrinsic spectrum of the source is simulated following a log-parabola shape with an integrated flux in the TeV band at the Crab level. A 50 h observation time is assumed with an energy resolution of 15 $\%$ and assuming a constant effective area of $10^5 \;\rm m^2$, these values being typical of current generation Cherenkov observatories. The intrinsic spectrum is convolved by one randomly generated photon survival probability and eventually binned to obtain the spectrum that would be observed in this model. The result of this simulation is displayed on the left panel of Fig.~\ref{fig:simu}. A fit of the simulated experimental data by a log-parabola shape convolved with EBL absorption is also shown, as it would be performed by observers. In the right panel of Fig.~\ref{fig:simu} are displayed the residuals of that fit. It appears that in the case without ALPs the residuals would evenly spread around 0 whereas these residuals would show anomalously strong and chaotic deviations from 0 in the case of $\gamma$-ALP mixing. This is the expected signature of ALPs in the spectrum, induced by the noisy spectral shape of the photon survival probability. 

The approach considered here corresponds to what an observer would do. First one would fit a smooth shape  (be it a log-parabola, a broken power  law or an emission model inspired one) and then pick up the shape providing the best $\chi^2$ and a decent residual distribution ({\it e.g.} centered on zero, without obvious biases, etc.). The crucial point is that the observer would fit the data with a smooth function. This is motivated by the fact that no TeV source emission model predicts spectra with local extrema or a noisy shape. So this result holds for any smooth spectral shape provided it gives the best possible fit. For a given observation, a given emission model and a given fit function, the noisy energy range and the variance of the residuals is a prediction of the ALP model. This is illustrated in Tab.~\ref{tab1}, which displays the variance of the fit residuals under different hypotheses: no ALP and two values of $g$, with an ALP mass still yielding $E_{\rm thr}=1\;\rm TeV$. The exact value of the variance of the fit residuals depends on the analysis that would be performed, in particular the energy range chosen by the observer. The use of the variance of the fit residuals is only an example, as observers might choose to use a more sophisticated estimator of the noise in theirs data sets. Additionally, because of the random nature of the predicted effect, it is important here to verify that the scatter in the prediction among realizations is smaller than the effect itself. The predicted uncertainty on the variance of the fit residuals due to the random nature of the prediction is also shown in Tab.~\ref{tab1} for each considered scenario. These results are obtained by averaging over 5000 realizations. It appears that for the considered parameters, the effect is significant, as one can see that the variance anomaly in the presence of ALP is predicted to be significantly above the conventional value. The observation of a Crab-level source for 50 h was chosen here as an illustration. Actually one finds that for the same redshift and energy range, the effect would still be visible but with less significance by observing only 5 h.

\begin{table}
\begin{tabular}{|c|c|}
\hline
\footnotesize
Model & Variance of the fit residuals \\ \hline
No ALP & $0.04\pm 0.01$ \\ \hline
$g=10^{-11}$, $m=0.7$ & $0.11\pm 0.04$ \\ \hline
$g=8\times10^{-11}$, $m=2$ & $0.20\pm 0.05$\\ \hline
\end{tabular}
\caption{Values of the RMS of the fit residuals to mock data with different assumptions for $g$ and $m$ (in units of  $\rm GeV^{-1}$ and neV resp.), for constant size magnetic field domains.\label{tab1}}
\end{table}

The observational signature that is discussed here occurs for energies around $E_{\rm thr}$ given in Eq.~\ref{thresh}. Therefore the range of accessible ALP parameters with this method depends on the value of the magnetic field and the energy range of the experiment. For instance considering TeV $\gamma$-rays and nG IGMF, the above results show that a typical IACT would be sensitive to ALPs with $g\sim10^{-11}\; \rm GeV^{-1}$ in a range from 0.1 neV to 10 neV. In that range of mass the most stringent constraint currently comes from the CAST helioscope with an upper limit on $g$ of order $10^{-10}\;\rm GeV^{-1}$~\cite{2011PhRvL.107z1302A}. So in principle this method should allow improving current constraints in this range of mass. To go to larger masses, one has to consider larger magnetic fields (in principle the method discussed in this article is valid for any $\gamma$-ray source behind a turbulent magnetic field) and/or higher energies, as the relevant mass for a given $g$ goes as $\sqrt{E\times B}$.

\section{Discussion}\label{discussion}

One important point is that should anomalous dispersion be observed some day, one would know how to 
falsify the interpretation in terms of new physics. This can be done for instance by observing the same object with more exposure. If the ALP interpretation is wrong, local extrema would not hold and all the residual points would be redistributed around zero. If the interpretation is correct though, two effects are predicted due to the increased statistics: {\it i}) the significance of the deviant bins would strengthen, and {\it ii}) irregularity would disappear at VHE as expected from ALP models. The first point is justified by the fact that a magnetic field that is coherent over a scale $s$ should remain coherent over times of oder $s/c$. For scales of order 1 Mpc as relevant here, this time scale is of order $3\times 10^6$ yrs. Concerning possible effects that could produce similar irregular spectra, one could imagine a complex landscape of background UV-IR photons that produces non-trivial absorption features in the energy spectra and mimics the effect. In the event  of a positive detection, this would therefore require studying the effect over more sources and how it depends on $z$ for instance. For observers interested in putting constraints on ALP models, though, this is not an issue since such an effect would add up to the irregularity of the spectrum and by no way it could cancel it.

The modeling of the IGMF as it is done here with domains of same sizes is the simplest model one could think of. It has been used here as it is widely used in the literature. To describe more precisely the magnetic field turbulence, it is possible to account for the power distribution of the modes. The turbulent field can be modeled as a Gaussian random field with each Fourier mode proportional to some power of the wave number $k^{-\alpha}$. In the generic case of isotropic and homogeneous Kolmogorov-like turbulence, $\alpha=5/3$. As shown in~\cite{2007PhRvD..76b3001M}, this leads to a variation of the rms intensity of the magnetic field $B$ as a function of the scale $s$ such that $B\propto s^{1/3}$.

Before discussing the effect of such a magnetic field on $\gamma$-ray source spectra, let us remark that magnetic fields that are coherent on small scales should have negligible effects on the spectrum in comparison with the larger scales. 
For small values of $\delta$, $P_0$, the conversion probability over a scale $s$, is expected to be of order $\delta^2/2\sim g^2B^2s^2/8$ (see Eq.~\ref{delta}). In that case, the averaged formula of Eq.~\ref{paveraged} reduces to $P_{\gamma\rightarrow a}\simeq N\delta^2/2$. All in all, for a given $g$, this probability is proportional to $N B^2 s^2$. If $P_s$ is the probability of photon conversion for modes of size $s$, given the above mentioned law for the magnetic field strength, one gets $P_{s/10}\sim 2.5\%\times P_s$ for the conversion probability in a magnetic field mode corresponding to a scale $s/10$. This means that the small scales rapidly become irrelevant for this study and one can safely consider that the largest scales contribute the most in the power distribution of modes. Concerning larger scales, the effect on the noise level is limited by the ratio between the considered scale and the distance to the source. Speaking in terms of domains, if there are only a few equivalent domains, little interference will happen and then the noise in the energy spectra will have wider fluctuations. 

\begin{table}
\begin{tabular}{|c|c|}
\hline
\footnotesize
Model & Variance of the fit residuals \\ \hline
$g=10^{-11}$, $m=0.35$ & $0.18\pm 0.05$ \\ \hline
$g=8\times10^{-11}$, $m=1$ & $0.42\pm 0.14$\\ \hline
\end{tabular}
\caption{Values of the RMS of the fit residuals to mock data with different assumptions for $g$ and $m$ (in units of  $\rm GeV^{-1}$ and neV resp.), in the case of a Kolmogorov-like turbulent magnetic field.\label{tab2}}
\end{table}

To be more quantitative, the study of Sec.~\ref{observational} has been repeated using a Kolmogorov-like turbulent magnetic field inspired by the modeling used in~\cite{1999ApJ...520..204G}. As for the previous study, 5000 realizations of turbulent magnetic field are performed, with wave numbers ranging from 0.1 Mpc to 100 Mpc, and a rms intensity of $B$ of 1 nG at 100 Mpc. The exact same kind of noise in the $\gamma$-ray spectra is obtained. To illustrate this, the results of these simulations are shown in Table ~\ref{tab2}; in particular, the variance of the fit residuals is still larger than in the no-ALP situation, in a statistically significant way. It has been checked that this results is stable when larger scales are used for the lowest wave number.

\section{Conclusion}

In this study we showed a new possible signature of $\gamma$-ALP mixing in the form of an anomalous dispersion in the energy spectra of $\gamma$-ray sources. The smooth-noisy-smooth alternation behavior in the energy spectrum is a peculiar prediction of ALP models that could hardly be mimicked by known astrophysical processes. It has been shown that this effect can be used to constrain ALP models from the observation of single sources. An explicit example has been given in the case of oscillations in IGMF; however, such a signature can be searched in any source for which a turbulent magnetic field is present along the line of sight.

\begin{acknowledgments}
We would like to thank Gilles Henri and Mathieu Langer for interesting discussions about the project,
and Pasquale Serpico, Jean-Fran\c{c}ois Glicenstein, Aion Viana, Emmanuel Moulin and Fabian Sch\"ussler for reading and improving the manuscript.
\end{acknowledgments}

\bibliography{wb}

\begin{thebibliography}{21}
\expandafter\ifx\csname natexlab\endcsname\relax\def\natexlab#1{#1}\fi
\expandafter\ifx\csname bibnamefont\endcsname\relax
  \def\bibnamefont#1{#1}\fi
\expandafter\ifx\csname bibfnamefont\endcsname\relax
  \def\bibfnamefont#1{#1}\fi
\expandafter\ifx\csname citenamefont\endcsname\relax
  \def\citenamefont#1{#1}\fi
\expandafter\ifx\csname url\endcsname\relax
  \def\url#1{\texttt{#1}}\fi
\expandafter\ifx\csname urlprefix\endcsname\relax\def\urlprefix{URL }\fi
\providecommand{\bibinfo}[2]{#2}
\providecommand{\eprint}[2][]{\url{#2}}

\bibitem[{\citenamefont{Peccei and Quinn}(1977{\natexlab{a}})}]{Peccei:1977hh}
\bibinfo{author}{\bibfnamefont{R.~D.} \bibnamefont{Peccei}} \bibnamefont{and}
  \bibinfo{author}{\bibfnamefont{H.~R.} \bibnamefont{Quinn}},
  \bibinfo{journal}{Phys. Rev. Lett.} \textbf{\bibinfo{volume}{38}},
  \bibinfo{pages}{1440} (\bibinfo{year}{1977}{\natexlab{a}}).

\bibitem[{\citenamefont{Peccei and Quinn}(1977{\natexlab{b}})}]{Peccei:1977ur}
\bibinfo{author}{\bibfnamefont{R.~D.} \bibnamefont{Peccei}} \bibnamefont{and}
  \bibinfo{author}{\bibfnamefont{H.~R.} \bibnamefont{Quinn}},
  \bibinfo{journal}{Phys. Rev.} \textbf{\bibinfo{volume}{D16}},
  \bibinfo{pages}{1791} (\bibinfo{year}{1977}{\natexlab{b}}).

\bibitem[{\citenamefont{{Sikivie}}(1983)}]{1983PhRvL..51.1415S}
\bibinfo{author}{\bibfnamefont{P.}~\bibnamefont{{Sikivie}}},
  \bibinfo{journal}{Physical Review Letters} \textbf{\bibinfo{volume}{51}},
  \bibinfo{pages}{1415} (\bibinfo{year}{1983}).

\bibitem[{\citenamefont{{Raffelt} and {Stodolsky}}(1988)}]{1988PhRvD..37.1237R}
\bibinfo{author}{\bibfnamefont{G.}~\bibnamefont{{Raffelt}}} \bibnamefont{and}
  \bibinfo{author}{\bibfnamefont{L.}~\bibnamefont{{Stodolsky}}},
  \bibinfo{journal}{\prd} \textbf{\bibinfo{volume}{37}}, \bibinfo{pages}{1237}
  (\bibinfo{year}{1988}).

\bibitem[{\citenamefont{{Arik} et~al.}(2011)\citenamefont{{Arik}, {Aune},
  {Barth}, {Belov}, {Borghi}, {Br{\"a}uninger}, {Cantatore}, {Carmona},
  {Cetin}, {Collar} et~al.}}]{2011PhRvL.107z1302A}
\bibinfo{author}{\bibfnamefont{M.}~\bibnamefont{{Arik}}},
  \bibinfo{author}{\bibfnamefont{S.}~\bibnamefont{{Aune}}},
  \bibinfo{author}{\bibfnamefont{K.}~\bibnamefont{{Barth}}},
  \bibinfo{author}{\bibfnamefont{A.}~\bibnamefont{{Belov}}},
  \bibinfo{author}{\bibfnamefont{S.}~\bibnamefont{{Borghi}}},
  \bibinfo{author}{\bibfnamefont{H.}~\bibnamefont{{Br{\"a}uninger}}},
  \bibinfo{author}{\bibfnamefont{G.}~\bibnamefont{{Cantatore}}},
  \bibinfo{author}{\bibfnamefont{J.~M.} \bibnamefont{{Carmona}}},
  \bibinfo{author}{\bibfnamefont{S.~A.} \bibnamefont{{Cetin}}},
  \bibinfo{author}{\bibfnamefont{J.~I.} \bibnamefont{{Collar}}},
  \bibnamefont{et~al.}, \bibinfo{journal}{Physical Review Letters}
  \textbf{\bibinfo{volume}{107}}, \bibinfo{eid}{261302} (\bibinfo{year}{2011}).

\bibitem[{\citenamefont{{Wagner} et~al.}(2010)\citenamefont{{Wagner}, {Rybka},
  {Hotz}, {Rosenberg}, {Asztalos}, {Carosi}, {Hagmann}, {Kinion}, {van Bibber},
  {Hoskins} et~al.}}]{2010PhRvL.105q1801W}
\bibinfo{author}{\bibfnamefont{A.}~\bibnamefont{{Wagner}}},
  \bibinfo{author}{\bibfnamefont{G.}~\bibnamefont{{Rybka}}},
  \bibinfo{author}{\bibfnamefont{M.}~\bibnamefont{{Hotz}}},
  \bibinfo{author}{\bibfnamefont{L.~J.} \bibnamefont{{Rosenberg}}},
  \bibinfo{author}{\bibfnamefont{S.~J.} \bibnamefont{{Asztalos}}},
  \bibinfo{author}{\bibfnamefont{G.}~\bibnamefont{{Carosi}}},
  \bibinfo{author}{\bibfnamefont{C.}~\bibnamefont{{Hagmann}}},
  \bibinfo{author}{\bibfnamefont{D.}~\bibnamefont{{Kinion}}},
  \bibinfo{author}{\bibfnamefont{K.}~\bibnamefont{{van Bibber}}},
  \bibinfo{author}{\bibfnamefont{J.}~\bibnamefont{{Hoskins}}},
  \bibnamefont{et~al.}, \bibinfo{journal}{Physical Review Letters}
  \textbf{\bibinfo{volume}{105}}, \bibinfo{eid}{171801} (\bibinfo{year}{2010}).

\bibitem[{\citenamefont{{Kim}}(1987)}]{1987PhR...150....1K}
\bibinfo{author}{\bibfnamefont{J.~E.} \bibnamefont{{Kim}}},
  \bibinfo{journal}{Phys. Rept.} \textbf{\bibinfo{volume}{150}},
  \bibinfo{pages}{1} (\bibinfo{year}{1987}).

\bibitem[{\citenamefont{{Raffelt}}(1996)}]{1996slfp.book.....R}
\bibinfo{author}{\bibfnamefont{G.~G.} \bibnamefont{{Raffelt}}},
  \emph{\bibinfo{title}{Stars as laboratories for fundamental\ physics : the
  astrophysics of neutrinos, axions, and other\ weakly interacting particles}}
  (\bibinfo{year}{1996}).

\bibitem[{\citenamefont{{Mirizzi} et~al.}(2007)\citenamefont{{Mirizzi},
  {Raffelt}, and {Serpico}}}]{2007PhRvD..76b3001M}
\bibinfo{author}{\bibfnamefont{A.}~\bibnamefont{{Mirizzi}}},
  \bibinfo{author}{\bibfnamefont{G.~G.} \bibnamefont{{Raffelt}}},
  \bibnamefont{and} \bibinfo{author}{\bibfnamefont{P.~D.}
  \bibnamefont{{Serpico}}}, \bibinfo{journal}{\prd}
  \textbf{\bibinfo{volume}{76}}, \bibinfo{eid}{023001} (\bibinfo{year}{2007}).

\bibitem[{\citenamefont{{Hooper} and {Serpico}}(2007)}]{2007PhRvL..99w1102H}
\bibinfo{author}{\bibfnamefont{D.}~\bibnamefont{{Hooper}}} \bibnamefont{and}
  \bibinfo{author}{\bibfnamefont{P.~D.} \bibnamefont{{Serpico}}},
  \bibinfo{journal}{Physical Review Letters} \textbf{\bibinfo{volume}{99}},
  \bibinfo{eid}{231102} (\bibinfo{year}{2007}).

\bibitem[{\citenamefont{{Hochmuth} and {Sigl}}(2007)}]{2007PhRvD..76l3011H}
\bibinfo{author}{\bibfnamefont{K.~A.} \bibnamefont{{Hochmuth}}}
  \bibnamefont{and} \bibinfo{author}{\bibfnamefont{G.}~\bibnamefont{{Sigl}}},
  \bibinfo{journal}{\prd} \textbf{\bibinfo{volume}{76}}, \bibinfo{eid}{123011}
  (\bibinfo{year}{2007}).

\bibitem[{\citenamefont{{De Angelis} et~al.}(2007)\citenamefont{{De Angelis},
  {Roncadelli}, and {Mansutti}}}]{2007PhRvD..76l1301D}
\bibinfo{author}{\bibfnamefont{A.}~\bibnamefont{{De Angelis}}},
  \bibinfo{author}{\bibfnamefont{M.}~\bibnamefont{{Roncadelli}}},
  \bibnamefont{and}
  \bibinfo{author}{\bibfnamefont{O.}~\bibnamefont{{Mansutti}}},
  \bibinfo{journal}{\prd} \textbf{\bibinfo{volume}{76}}, \bibinfo{eid}{121301}
  (\bibinfo{year}{2007}).

\bibitem[{\citenamefont{{S{\'a}nchez-Conde}
  et~al.}(2009)\citenamefont{{S{\'a}nchez-Conde}, {Paneque}, {Bloom}, {Prada},
  and {Dom{\'{\i}}nguez}}}]{2009PhRvD..79l3511S}
\bibinfo{author}{\bibfnamefont{M.~A.} \bibnamefont{{S{\'a}nchez-Conde}}},
  \bibinfo{author}{\bibfnamefont{D.}~\bibnamefont{{Paneque}}},
  \bibinfo{author}{\bibfnamefont{E.}~\bibnamefont{{Bloom}}},
  \bibinfo{author}{\bibfnamefont{F.}~\bibnamefont{{Prada}}}, \bibnamefont{and}
  \bibinfo{author}{\bibfnamefont{A.}~\bibnamefont{{Dom{\'{\i}}nguez}}},
  \bibinfo{journal}{\prd} \textbf{\bibinfo{volume}{79}}, \bibinfo{eid}{123511}
  (\bibinfo{year}{2009}).

\bibitem[{\citenamefont{{Tavecchio} et~al.}(2012)\citenamefont{{Tavecchio},
  {Roncadelli}, {Galanti}, and {Bonnoli}}}]{2012arXiv1202.6529T}
\bibinfo{author}{\bibfnamefont{F.}~\bibnamefont{{Tavecchio}}},
  \bibinfo{author}{\bibfnamefont{M.}~\bibnamefont{{Roncadelli}}},
  \bibinfo{author}{\bibfnamefont{G.}~\bibnamefont{{Galanti}}},
  \bibnamefont{and}
  \bibinfo{author}{\bibfnamefont{G.}~\bibnamefont{{Bonnoli}}},
  \bibinfo{journal}{ArXiv e-prints}  (\bibinfo{year}{2012}).

\bibitem[{\citenamefont{{Mirizzi} and {Montanino}}(2009)}]{2009JCAP...12..004M}
\bibinfo{author}{\bibfnamefont{A.}~\bibnamefont{{Mirizzi}}} \bibnamefont{and}
  \bibinfo{author}{\bibfnamefont{D.}~\bibnamefont{{Montanino}}},
  \bibinfo{journal}{JCAP} \textbf{\bibinfo{volume}{12}}, \bibinfo{pages}{4}
  (\bibinfo{year}{2009}).

\bibitem[{\citenamefont{{Burrage} et~al.}(2009)\citenamefont{{Burrage},
  {Davis}, and {Shaw}}}]{2009PhRvL.102t1101B}
\bibinfo{author}{\bibfnamefont{C.}~\bibnamefont{{Burrage}}},
  \bibinfo{author}{\bibfnamefont{A.-C.} \bibnamefont{{Davis}}},
  \bibnamefont{and} \bibinfo{author}{\bibfnamefont{D.~J.}
  \bibnamefont{{Shaw}}}, \bibinfo{journal}{Physical Review Letters}
  \textbf{\bibinfo{volume}{102}}, \bibinfo{eid}{201101} (\bibinfo{year}{2009}).

\bibitem[{\citenamefont{Horns and Meyer}(2012)}]{Horns:2012fx}
\bibinfo{author}{\bibfnamefont{D.}~\bibnamefont{Horns}} \bibnamefont{and}
  \bibinfo{author}{\bibfnamefont{M.}~\bibnamefont{Meyer}},
  \bibinfo{journal}{JCAP} \textbf{\bibinfo{volume}{1202}}, \bibinfo{pages}{033}
  (\bibinfo{year}{2012}), \eprint{1201.4711}.

\bibitem[{\citenamefont{{Cs{\'a}ki} et~al.}(2003)\citenamefont{{Cs{\'a}ki},
  {Kaloper}, {Peloso}, and {Terning}}}]{2003JCAP...05..005C}
\bibinfo{author}{\bibfnamefont{C.}~\bibnamefont{{Cs{\'a}ki}}},
  \bibinfo{author}{\bibfnamefont{N.}~\bibnamefont{{Kaloper}}},
  \bibinfo{author}{\bibfnamefont{M.}~\bibnamefont{{Peloso}}}, \bibnamefont{and}
  \bibinfo{author}{\bibfnamefont{J.}~\bibnamefont{{Terning}}},
  \bibinfo{journal}{JCAP} \textbf{\bibinfo{volume}{5}}, \bibinfo{pages}{5}
  (\bibinfo{year}{2003}).

\bibitem[{\citenamefont{{Kneiske} and {Dole}}(2010)}]{2010A&A...515A..19K}
\bibinfo{author}{\bibfnamefont{T.~M.} \bibnamefont{{Kneiske}}}
  \bibnamefont{and} \bibinfo{author}{\bibfnamefont{H.}~\bibnamefont{{Dole}}},
  \textbf{\bibinfo{volume}{515}}, \bibinfo{eid}{A19} (\bibinfo{year}{2010}).

\bibitem[{\citenamefont{{Grossman} et~al.}(2002)\citenamefont{{Grossman},
  {Roy}, and {Zupan}}}]{2002PhLB..543...23G}
\bibinfo{author}{\bibfnamefont{Y.}~\bibnamefont{{Grossman}}},
  \bibinfo{author}{\bibfnamefont{S.}~\bibnamefont{{Roy}}}, \bibnamefont{and}
  \bibinfo{author}{\bibfnamefont{J.}~\bibnamefont{{Zupan}}},
  \bibinfo{journal}{Physics Letters B} \textbf{\bibinfo{volume}{543}},
  \bibinfo{pages}{23} (\bibinfo{year}{2002}).

\bibitem[{\citenamefont{{Giacalone} and {Jokipii}}(1999)}]{1999ApJ...520..204G}
\bibinfo{author}{\bibfnamefont{J.}~\bibnamefont{{Giacalone}}} \bibnamefont{and}
  \bibinfo{author}{\bibfnamefont{J.~R.} \bibnamefont{{Jokipii}}},
  \bibinfo{journal}{\apj} \textbf{\bibinfo{volume}{520}}, \bibinfo{pages}{204}
  (\bibinfo{year}{1999}).

\end{thebibliography}

\end{document}